# CAN LARGE LANGUAGE MODELS FIND AND FIX VULNERABLE SOFTWARE?

David Noever[1]
PeopleTec, 4901-D Corporate Drive, Huntsville, AL, USA, 35805
[1]david.noever@peopletec.com

## ABSTRACT

*In this study, we evaluated the capability of Large Language Models (LLMs), particularly OpenAI's GPT-4, in detecting software vulnerabilities, comparing their performance against traditional static code analyzers like Snyk and Fortify. Our analysis covered numerous repositories, including those from NASA and the Department of Defense. GPT-4 identified approximately four times the vulnerabilities than its counterparts. Furthermore, it provided viable fixes for each vulnerability, demonstrating a low rate of false positives. Our tests encompassed 129 code samples across eight programming languages, revealing the highest vulnerabilities in PHP and JavaScript. GPT-4's code corrections led to a 90% reduction in vulnerabilities, requiring only an 11% increase in code lines. A critical insight was LLMs' ability to self-audit, suggesting fixes for their identified vulnerabilities and underscoring their precision. Future research should explore system-level vulnerabilities and integrate multiple static code analyzers for a holistic perspective on LLMs' potential.*

## KEYWORDS

*ChatGPT, human, response, metrics, vulnerabilities, software fix, code languages*

## 1. INTRODUCTION

Software systems' increasing complexity and ubiquity demands advanced methods for ensuring their security. While traditional rule-based static code analyzers, such as HP Fortify or Snyk, have been instrumental in identifying software vulnerabilities, their rule-based nature can sometimes miss nuanced or evolving threats [1-9]. Large Language Models (LLMs) like OpenAI's ChatGPT [10-29] offer a novel avenue for addressing this challenge. Powered by vast amounts of textual data, LLMs have shown potential in understanding and generating code, suggesting they could be adept at pinpointing and rectifying software vulnerabilities [10,16].

Recent explorations have showcased the capabilities of LLMs in detecting security vulnerabilities, sometimes even outperforming traditional methods [2,3]. For instance, an LLM identified 213 security vulnerabilities in a single codebase, underscoring its potential as a security tool[2]. Moreover, LLMs have been applied to various coding contexts, from visual programming to Java functions, often revealing proficiency in code generation and understanding [18,21]. Furthermore, LLMs demonstrate adaptability, tackling challenges like code evolution, high-performance computing, and self-collaboration code generation [25,26,27].

In this paper, we take a comprehensive look at the performance of LLMs in identifying and rectifying software vulnerabilities. By examining numerous repositories on GitHub using both LLMs and code analyzers, we aim to contrast their efficacy in addressing software security concerns. This comparison will provide insights into LLMs' strengths and limitations in software vulnerability detection and rectification. Table 1 compares the two approaches in their intent and design to recognize software vulnerabilities.

While some evaluations indicate the potential of ChatGPT in vulnerability detection, the broader implications of LLMs in software security remain an open question [28]. As the software landscape continues to evolve, leveraging the capabilities of LLMs in tandem with traditional methods might pave the way for more secure and robust systems. This study aims to contribute to this evolving discourse, providing a clearer picture of where LLMs stand in the quest for software security. Because previous work [20-27] focuses on LLMs that generate code from text descriptions, the present work extends the vulnerability recognition work [2-3, 28] to include mitigations and software fixes.

We extend the test datasets from traditional programming categories like buffer overflow or command injections to analyze public and important scientific repositories [30-36] from NASA Flight Systems [30] and Code Analyzer [36], the National Geospatial-Intelligence Agency [31], Department of Defense challenges [32] and Android Tactical Assault Kit (ATAK) [33], and leading AI vision [34] and Microsoft Research's cyber agents and reinforcement libraries [35]. By generalizing the challenges beyond just programming case studies, we seek to make contact [29] with the scientific community that uses collaborative coding tools to augment their skills but often does not seek to become professional software developers themselves. This demographic benefits from a personal coding assistant to directly stimulate new ideas for addressing problems [9,26-27, 29] in math, physics, chemistry, biology, and software engineering [16].

**Table 1. Overview of the core differences and similarities between the two approaches to code analysis**

| Aspect | Static Code Analyzers (e.g., Snyk, Fortify) | Large Language Models (LLMs) |
| --- | --- | --- |
| **Purpose and Design** | Designed to identify known security vulnerabilities in code | Designed to understand and generate human-like text, including code |
| **Code Representation** | Use Abstract Syntax Trees (ASTs) or Control Flow Graphs (CFGs) | Represent code as sequences of tokens |
| **Learning and Adaptation** | Rely on predefined rules and signatures; don't traditionally "learn" | Continuously learn from training data; adapt based on seen patterns |
| **Generalization** | Precise and specific; based on known patterns/signatures | Can generalize across various coding patterns/styles |
| **Feedback and Iteration** | Deterministic feedback based on rule matching | Provide contextual, descriptive feedback |
| **Coverage** | Limited to set of predefined rules/signatures | Potentially broader due to generalized training, but may lack pinpoint accuracy |
| **Basis of Operation** | Rule-based | Pattern recognition based on training data |
| **Adaptability** | Fixed unless rules are updated | Flexible due to pattern recognition capabilities |

| Primary Use Case | Security vulnerability detection | Text understanding, generation, and contextual reasoning |

As a final evaluation, we ask GPT-4 to rewrite the vulnerable code removing all the issues, and then score the same repository post-correction. This approach presents a unique test for a LLM to act not just as a code scanner but also as an auditable fixer of vulnerabilities in code corrections.

## 2. METHODS

The study includes the latest Open AI models (GPT-4) accessed automatically through the chat interface with a system context set to *"act as the world's greatest static code analyzer for all major programming languages. I will give you a code snippet, and you will identify the language and analyze it for vulnerabilities. Give the output in a format: filename, vulnerabilities detected as a numbered list, and proposed fixes as a separate numbered list."*

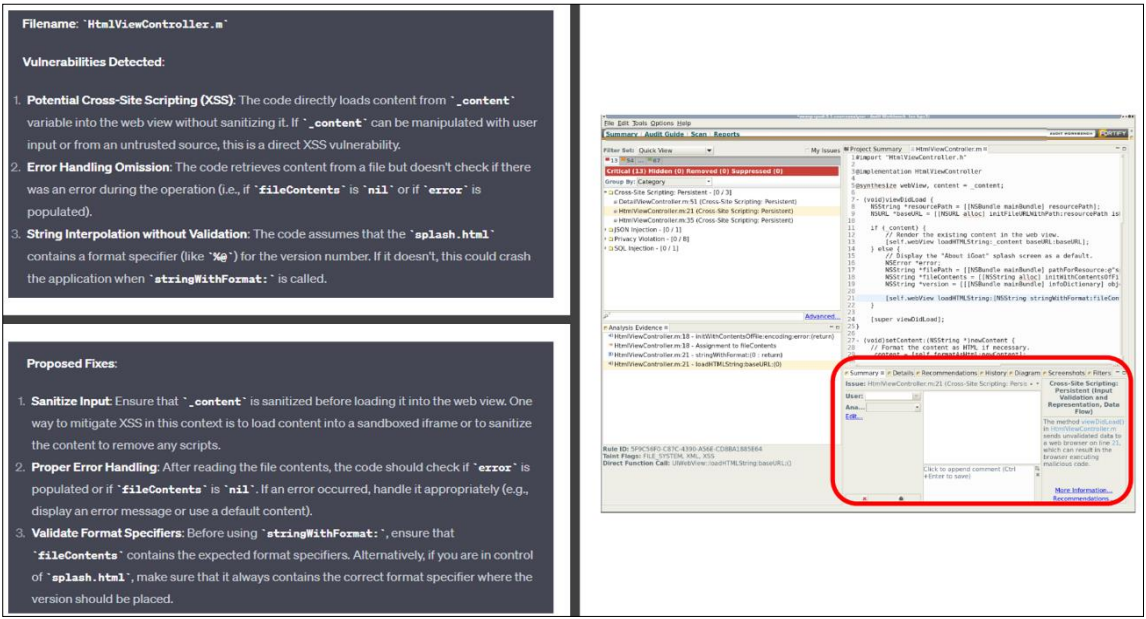

**Figure 1. Comparison between LLM (left-GPT-4) and static code analyzer (right-HP Fortify) show Cross-Site Scripting (XSS) in an Objective-C method called HtmlViewController.**

We apply this context to 7 different LLMs from OpenAI ranging in parameter sizes [37] that span four orders of magnitude: 350M (Ada), 6.7B (Curie), 175B (DaVinci/ GPT3/GPT-3.5-turbo-16k), and 1.7 trillion (GPT-4). However, the details are proprietary, at least the larger models, since Curie has trained on gigabytes of GitHub software repositories in multiple languages and the OpenAI Codex source codes. In practice, scaling above 13 billion parameters offers the first hint of programmer skills for more than code commentary or auto-completion based on memorization in training data. Since the [2] publication in February 2023, multiple GPT version updates have demonstrated significant advances from larger models (by orders of magnitude in parameter scaling). OpenAI grades the coding performance improvements [10] across a range of skills, including (easy) Leetcode exam scores that grew from 12/41 correct (GPT-3.5) to 31/41 right (GPT-4).

In all cases, we query the LLMs automatically using the API, a system context to look for vulnerabilities and fixes, followed by sample code in eight popular programming languages (C, Ruby, PHP, Java, Javascript, C#, Go, and Python). In each case, we ask the LLM to identify the coding language, find vulnerabilities and propose fixes.

The Single Codebase of Security Vulnerabilities [2] includes 128 code snippets with examples in all eight programming languages that illustrate thirty-three vulnerable categories. The categories range from Buffer Overflow to Sensitive Data Exposure. Fifty cases involve PHP with vulnerabilities related to file inclusion and command injections. The vulnerability-language matrix illustrates Appendix A's total coverage for exploring code scanners. The entire software lines of executable code equals 2372, excluding markdown and HTML.

We submitted six public repositories [30-36] on GitHub to the automated static code scanner, Snyk [1], a project relied upon by millions of developers and subscribers listed from Amazon AWS Cloud, Google, Salesforce, Atlassian, and Twilio. The role of these diverse scans was to illustrate the plethora of identifiable vulnerabilities ("finding") and the breadth of language problems addressed by LLMs. For each file, Snyk offers its vulnerability intelligence dashboard with comprehensive metrics for Severity (Critical, High, Medium, Low), Priority Score (0-1000), Fixability (Fixable, Partially fixable, No fix available), Exploit Maturity (Mature, Proof of Concept, No Known Exploit, and No Data), Status (Open, Patched, Ignored), and Dependency Issues.

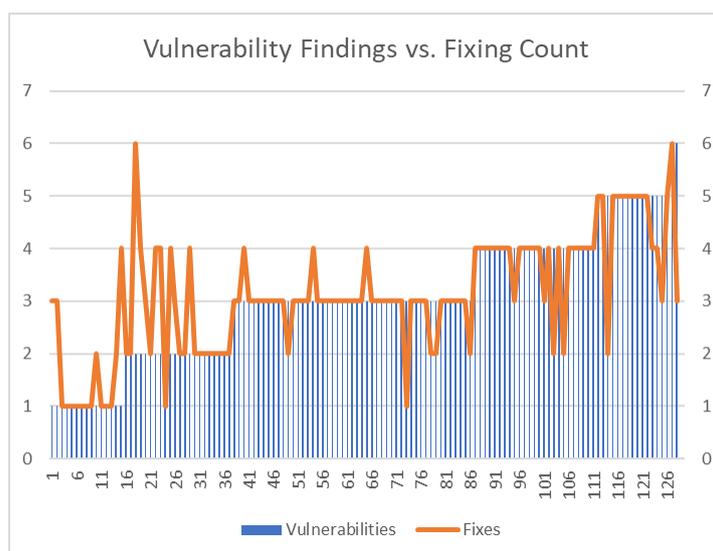

Figure 2. Sorted Count of Vulnerabilities Found and Fixed Show Strong Relationship.

As a final evaluation, we submit all 129 code samples [2] to GPT-4 automatically using the API, but this time the prompt seeks corrected code only. The general system prompt reads: *"Act as the world's greatest static code analyzer for all major programming languages. I will give you a code snippet, and you will analyze the code and rewrite it, removing any identified vulnerabilities. Do not explain, just return the corrected code and format alone."*

After post-processing the returned code from GPT-4 into files and uploading the files to a scannable GitHub, we resubmit the corrected repository to Snyk to compare against the vulnerable codebase. In this way, the evaluation seeks to score the self-correction capabilities of LLMs as automating not just the identification of vulnerabilities but the rewriting of code to secure the entire codebase as objectively validated by a third-party static code scanner.

| Vulnerability Category | LLM Vulnerabilities | LLM Fixes | Snyk Vulnerabilities |
|---|---|---|---|
| Path Traversal | 46 | 46 | 16 |
| File Inclusion | 40 | 45 | 12 |
| Command Injection | 34 | 43 | 13 |
| SQL Injection | 30 | 26 | 6 |
| Unsafe Deserialization | 25 | 23 | 2 |
| Insecure File Uploads | 19 | 19 | |
| PHP Object Injection | 18 | 18 | |
| Cross-site Scripting (XSS) | 17 | 18 | 11 |
| Buffer Overflow | 16 | 18 | |
| Denial Of Service | 14 | 14 | 5 |
| Server Side Template Injection | 11 | 11 | |
| Connection String Injection | 11 | 11 | |
| XML External Entity (XXE) Injection | 11 | 14 | 3 |
| PostMessage Security | 10 | 9 | |
| Code Injection | 9 | 10 | 1 |
| LDAP Injection | 9 | 9 | |
| Sensitive Data Exposure | 6 | 4 | 1 |
| Open Redirect | 6 | 7 | 1 |
| Format String Attacks | 6 | 3 | |
| Use After Free | 5 | 3 | |
| Integer Overflow | 5 | 5 | |
| Out of Bounds | 5 | 5 | |
| Zip Traversal | 4 | 4 | |
| IDOR | 4 | 3 | |
| Server-Side Request Forgery (SSRF) | 4 | 4 | 3 |
| Symlink Attack | 4 | 4 | |
| ReDoS | 4 | 4 | |
| Code Execution | 3 | 3 | |
| Resource Injection | 3 | 3 | |
| Prototype Pollution | 3 | 1 | |
| NoSQL Injection | 3 | 3 | |
| XPATH Injection | 2 | 2 | 1 |
| Log Forging | 1 | 1 | 1 |
| Other | | | 22 |
| Grand Total | 393 | 398 | 98 |

**Figure 3. Comparison by Vulnerability Categories for LLM vs Snyk**

## 3. RESULTS

An initial example [3] to illustrate the comparison between static code analyzer (HP Fortify) results and an LLM (OpenAI's GPT-4 2023AUG3 version) is shown in Figure 1. Both approaches correctly identify the three vulnerabilities in an Objective-C method called *HtmlViewController.m* file. The LLM version, however, offers a plain English explanation of why the vulnerability arises and how an attacker might exploit it with user inputs. However, the significant difference in this example is the proposed three fixes for each vulnerability. To complete the vulnerability finding and fixing process, the LLM offers revised code that patches each of the three identified cases to sanitize the input, check for errors after file reads, and validate the expected string format.

The published list of vulnerabilities presented by the Single Codebase of Security Vulnerabilities [2] author noted that GPT-3 (text-davinci-003) found 213 security issues compared to Snyk's 99 (excluding the 16 files in C and Go, which Snyk does not cover). In the random manual inspection of 60 of 213 vulnerabilities, only four were false positives, but both tools (DaVinci) and Snyk had many false negatives or missed vulnerabilities.

The present work scanned the same codebase with Snyk and identified 98 vulnerabilities in our runs, with approximately two-thirds of the vulnerabilities ranked high severity (H-66, M-20, L-12). Using the 3AUG2023 GPT-4 API, our results show 393 identified vulnerabilities, almost twice as many as DaVinci (213) and four times the number found by Snyk (99). Our scoring of GPT-3-Turbo-16K roughly corresponds to [2] results for GPT-3, with an identified vulnerability count of 217. Random inspection of the Curie and Ada models shows the degenerate repetition of industry jargon and no concrete proposals to fix vulnerabilities. This observation suggests that somewhere between 6 billion and 175 billion parameters, significant code understanding emerges in the OpenAI GPT series.

One feature of interest is the number of proposed code fixes also equals 398 for GPT-4, which supports a low false positive rate since asking for a solution forces the model to justify the identification of the vulnerability and correct any misstated or hallucinatory responses. Figure 2 shows the connection between vulnerabilities found and patches proposed when sorted by the number of vulnerabilities in each of the 129 files. The chart bolsters the idea that a true positive must also have a proposed fix (or, in some cases, more than one).

| Table 2. Single Codebase of Security Vulnerabilities (SLOC = SW Lines of Code) | | | | | |
|---|---|---|---|---|---|
| **Codebase & Reference** | **SLOC** | **Critical** | **High** | **Med** | **Low** |
| Original GitHub Repo [2] | 2372 | 0 | 66 | 20 | 12 |
| GPT-4 Corrected GitHub Repo [38] | 2636 | 0 | 4 | 5 | 1 |
| **Difference** | **+264** | **0** | **-94%** | **-75%** | **-92%** |

Figure 3 highlights the vulnerability categories LLM analysis (GPT-4) found compared to Snyk. For the top two vulnerability categories (Path Traversal and File Inclusion), GPT-4 identified three to four times as many security flaws and similarly proposed a fix for each finding.

Table 2 summarizes the evaluation results for automating LLM code corrections. Compared to the original vulnerable codebase, the LLM added 11% more software lines of code to mitigate its identified vulnerabilities. When Snyk scores the severity of vulnerabilities before and after corrections, the LLM reduced the number of high-severity vulnerabilities by 94%, medium by 75%, and low by 92%. In absolute numbers, the codebase mitigations lower the vulnerabilities from 98 to 10.

Figure 4 shows the LLM identification of vulnerabilities and fixes by programming language, with PHP and JavaScript accounting for nearly half of the findings. GPT-4 succeeded in all programming languages to correctly identify the type (e.g., python) without including that hint in the system context. This automated code identification serves potential automation goals for larger code bases with a complicated legacy software stack.

Appendix B highlights a side-by-side comparison of the Snyk output vs. the LLM (GPT-4) analysis for a vulnerable image uploader written in PHP.

To explore more prominent, complicated codebases of real-world significance, we submitted seven public GitHub repositories to Snyk and collected the number and severity of identified vulnerabilities in Table 3. The Hack a Satellite (HAS) project in the Department of Defense repository refers to qualifying challenges before a DEFCON hackathon, thus emphasizing some potential vulnerabilities as part of its mission. The NASA Software to perform software scanning for validation and verification (VnV) shows the second-highest vulnerability count as scanned by Snyk.

| Table 3. Large Codebases and Snyk Identified Severe Software Vulnerabilities | | | | | | |
|---|---|---|---|---|---|---|
| Codebase & Reference | Stars | SLOC | Critical | High | Med | Low |
| DoD Hack a Satellite [32] | 92 | 5.73M | 59 | 209 | 274 | 4000 |
| NASA Software VnV -ikos [36] | 1800 | 44k | 4 | 58 | 999 | 1300 |
| DoD Tactical Assault Kit [33] | 650 | 1.44M | | 140 | 863 | 55 |
| NGA SarPy Radar Satellite [31] | 195 | 144k | | 1 | 15 | 10 |
| Ultralytics YOLOv5 [34] | 40900 | 15k | | | 55 | 7 |
| Microsoft CyberBattleSim [35] | 1500 | 8k | | | 3 | 6 |

The table shows GitHub stars as a proxy for its popularity and, thus, its potential for exploitation in the wild. We used the cloc executable to calculate the software lines of code (SLOC) as a proxy measure of code complexity.

The most popular repository in this selection is the object detection library by Ultralytics [34], used by many computer vision projects. In 2022, the most popular repository (TensorFlow) totalled 177,000 stars, followed by Linux with 156,000 stars. In this relative hierarchy, one might infer that YOLOv5 rivals some of the largest public software bases in popularity with 40,900 stars. YOLOv5 totals only four times more SLOC than the vulnerable codebase [2], suggesting a roughly similar code complexity, although much of the Python code relies on unscanned library dependencies such as open-cv. Beyond proxies for complexity and popularity, the vulnerabilities in NASA software scanners warrant further investigation, given this repo serves as an authorizing agent for what might prove to be critical space assets or endanger human life if flawed in some uncorrected ways.

Examples of rule-based code scanners and LLMs for these projects are shown in Appendix B.

## 4. DISCUSSION

One motivation for the present study was to extend the Single Codebase of Security Vulnerabilities' benchmark to identify the false negatives and to catalog the fixes and distribution across coding languages and vulnerability classes. One hypothesis to test here is whether the larger quantity of some languages like Python, C, and Java on GitHub benefits the models' ability to find vulnerable code.

A second motivation centers on the LLM criticism that their output appears unreliable or prone to hallucination since its underlying optimization is to perform the next token or word prediction, and only later in its maturity does it benefit from reliability checks. OpenAI concedes on its chat interface, "ChatGPT may produce inaccurate information about people, places, or facts."

| Languages | Vulnerabilities | Fixes |
|---|---|---|
| php | 159 | 170 |
| js | 63 | 57 |
| cs | 50 | 58 |
| c | 49 | 46 |
| java | 27 | 26 |
| py | 19 | 18 |
| rb | 11 | 10 |
| other (html, etc) | 6 | 6 |
| ts | 5 | 3 |
| go | 4 | 4 |
| Grand Total | 393 | 398 |

**Figure 4. GPT-4 Findings and Fixes by Programming Language**

Therefore, this work investigates whether matching vulnerabilities with fixes is a self-reflective trigger, forcing the model to re-evaluate its initial conclusions. In a sense, the most valid test of a model's reliability in identifying flaws is to force it to find the fix, then implement the fixes in code that works. That virtuous cycle supports a strong case for supplementing rule-based expert systems like static code analyzers with LLMs. If one assumes the GPT-4 case is reasonably close to the actual number of vulnerabilities, this finding suggests that between 175 billion and 1.7 trillion parameters, the LLM series reduces false negatives by half compared to smaller models and by three-quarters compared to the Snyk instance. A future initiative should extend this analysis to other static code analyzers beyond HP Fortify and Snyk for generality; SonarQube remains a popular open-source version to try.

It is worth noting that the code snippets tested are typically a few tens or hundreds of lines long and thus limited below about 500 tokens sent to the model. This prompt and a token response limit (1024) prevent the API approach from analyzing system-level vulnerabilities or looking at dependencies or library errors. Because of the size of the vulnerability test, this approach makes sense, but future work using OpenAI's recently released Code Interpreter [39] should enable system-level inquiries, as illustrated in Appendix B.

The most notable results include the four-fold increase in vulnerabilities found using LLM as a code scanner, followed by a 90% reduction in vulnerabilities using GPT-4 code corrections. In a parsimonious way, only 11% of additional software lines of code remove the identified security flaws.

Without the API currently, follow-up questions from a manual user could likely expand the depth of model answers given the right prompt and hint to explore in further analysis. One intriguing but often underestimated advantage of "asking questions" to one's code is the interactive format stimulates ideas that a static response tends to flatten or transform into toil and a long to-do list. One can anticipate that the future copilot or coding assistant role will evolve into a mentor-protégé relationship where software requirements are prioritized between humans and machines for the most efficient and least vulnerable software delivered.

## ACKNOWLEDGMENTS

The author benefited from the encouragement and project assistance of the PeopleTec Technical Fellows program. The author thanks the researchers at OpenAI for developing large language models and allowing public access to ChatGPT.## REFERENCES

**Authors**

**David Noever** has research experience with NASA and the Department of Defense in machine learning and data mining. He received his BS from Princeton University and his Ph.D. from Oxford University, as a Rhodes Scholar, in theoretical physics.

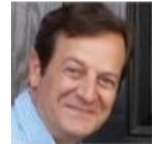

**Appendix A: Vulnerability-Language Matrix of Find and Fix Coverage in Single Codebase Dataset [2]**

| Vulnerability Category | asp | aspx.cs | c | cs | go | html | java | js | php | py | rb | ts | Total |
|---|---|---|---|---|---|---|---|---|---|---|---|---|---|
| Buffer Overflow | | | 6 | | | | | | | | | | 6 |
| Code Execution | | | | | | | | | | | 1 | | 1 |
| Code Injection | | | | | | | | | 2 | | 1 | | 3 |
| Command Injection | | | | 1 | | | | 1 | 9 | 1 | | | 12 |
| Connection String Injection | 1 | | 1 | | | | 1 | | | | | | 3 |
| Denial Of Service | | | 1 | 1 | | | | 1 | | | | | 3 |
| File Inclusion | | | | | | | | | 14 | | | | 14 |
| Format String Attacks | | | 2 | | | | | | | | | | 2 |
| IDOR | | | | | | | | | 1 | | | | 1 |
| Insecure File Uploads | | | | | | | | | 5 | | | | 5 |
| Integer Overflow | | | 1 | | | | | | | | | | 1 |
| LDAP Injection | 1 | | | 1 | | | | | 1 | | | | 3 |
| Log Forging | | | | 1 | | | | | | | | | 1 |
| NoSQL Injection | | | | | | | | 1 | | | | | 1 |
| Open Redirect | | | | | | | | 2 | 1 | | 1 | | 4 |
| Out of Bounds | | | 1 | | | | | | | | | | 1 |
| Path Traversal | | | | 4 | | | 4 | 2 | 3 | 1 | | | 14 |
| PHP Object Injection | | | | | | | | | 5 | | | | 5 |
| PostMessage Security | | | | | | | | 3 | | | | | 3 |
| Prototype Pollution | | | | | | | | 1 | | | | | 1 |
| ReDoS | | | | | | | | 1 | | | | | 1 |
| Resource Injection | 1 | | | | | | | | | | | | 1 |
| Sensitive Data Exposure | | | | | | | | | | | 1 | 2 | 3 |
| Server Side Template Injection | | | | | | 1 | | | 1 | 2 | | | 4 |
| SQL Injection | | | | 1 | | | 1 | 3 | 3 | | 1 | | 9 |
| SSRF | | | | | | | | 1 | | | | | 1 |
| Symlink Attack | | | 1 | | | | | | | | | | 1 |
| Unsafe Deserialization | | | | | | | 2 | 2 | 1 | 2 | | | 7 |
| Use After Free | | | 1 | | | | | | | | | | 1 |
| XPATH Injection | | | | 1 | | | | | | | | | 1 |
| XSS | | | | 3 | | 1 | 1 | 1 | 2 | | | | 8 |
| XXE | | | | 2 | | | | | 1 | 2 | | | 5 |
| Zip Traversal | | | | 1 | | | | | | | | | 1 |
| Total | 3 | 1 | 14 | 15 | 1 | 1 | 9 | 20 | 50 | 6 | 5 | 2 | 128 |

**Appendix B: Example Vulnerabilities and Fixes Found by GPT-4 vs. Snyk on Security Codebase**

| Code: Insomnihack_2019_l33t-hoster.php | GPT-4 and Snyk Findings |
|---|---|
| ```<br><?php<br>if (isset($_GET["source"]))<br>    die(highlight_file(__FILE__));<br>session_start();<br>if (!isset($_SESSION["home"])) {<br>    $_SESSION["home"] = bin2hex(random_bytes(20));<br>}<br>$userdir = "images/{$_SESSION["home"]}/";<br>if (!file_exists($userdir) {<br>``` | **GPT-4 Analysis**<br><br>The code is written in PHP and it's a script for uploading images.<br><br>Filename: upload.php<br><br>Vulnerabilities detected:<br><br>1. The script allows for Path Traversal, it uses the session identifier to create a directory where it stores the uploaded files. This could potentially be exploited |
| ```<br>    mkdir($userdir);<br>}<br>$disallowed_ext = array(<br>    "php",<br>    "php3",<br>    "php4",<br>    "php5",<br>    "php7",<br>    "pht",<br>    "phtm",<br>    "phtml",<br>    "phar",<br>    "phps",<br>);<br>if (isset($_POST["upload"])) {<br>    if ($_FILES['image']['error'] !== UPLOAD_ERR_OK) {<br>        die("yuuuge fail");<br>    }<br>    $tmp_name = $_FILES["image"]["tmp_name"];<br>    $name = $_FILES["image"]["name"];<br>    $parts = explode(".", $name);<br>    $ext = array_pop($parts);<br>    if (empty($parts[0])) {<br>        array_shift($parts);<br>    }<br>    if (count($parts) === 0) {<br>        die("Filename is empty");<br>    }<br>    if (in_array($ext, $disallowed_ext, TRUE)) {<br>        die("Lorem ipsum dolor sit amet consectetur adipisicing elit.");<br>    }<br>    $image = file_get_contents($tmp_name);<br>    if (mb_strpos($image, "<?") !== FALSE) {<br>        die("Lorem ipsum dolor sit amet consectetur adipisicing elit.");<br>    }<br>    if (!exif_imagetype($tmp_name)) {<br>        die("Lorem ipsum dolor sit amet consectetur adipisicing elit.");<br>    }<br>    $image_size = getimagesize($tmp_name);<br>    if ($image_size[0] !== 1337  $image_size[1] !== 1337) {<br>``` | if the session identifier is predictable or can be controlled by an attacker.<br><br>2. The script does not validate the file name of the uploaded file, which could lead to an arbitrary file upload vulnerability. This could be exploited to upload a malicious file.<br><br>3. The script uses `die()` function for error handling, which can expose sensitive information to the user or attacker.<br><br>4. The script does not sanitize the output when displaying the file list, leading to a potential Cross-Site Scripting (XSS) vulnerability.<br><br>Proposed fixes:<br><br>1. Use a secure method to generate the directory name, and ensure it cannot be controlled or predicted by an attacker.<br><br>2. Validate the file name of the uploaded file to ensure it only contains allowed characters.<br><br>3. Use a custom error handler that does not expose sensitive information.<br><br>4. Sanitize the output when displaying the file list to prevent XSS attacks.<br><br>______________________<br><br>**Snyk Analysis**<br><br>Unsanitized input from an uploaded file flows into move_uploaded_file, where it is used as a path. This may result in a Path Traversal vulnerability and allow an attacker to move arbitrary files. |

```
        die("Lorem ipsum dolor sit amet consectetur
adipisicing elit.");
    }
    $name = implode(".", $parts);
    move_uploaded_file($tmp_name, $userdir . $name .
"." . $ext);
}
echo "<h3>Your <a href=$userdir>files</a>:</h3><ul>";
foreach(glob($userdir . "*") as $file) {
    echo "<li><a href='$file'>$file</a></li>";
}
echo "</ul>";
?>
<h1>Upload your pics!</h1>
<form method="POST" action="?"
enctype="multipart/form-data">
    <input type="file" name="image">
    <input type="submit" name=upload>
</form>
```

| **analyzer/python/ikos/scan.py**<br><br>```<br>    ''' Run the given command and return the<br>exit code '''<br>    log.debug('Running %s' %<br>command_string(cmd))<br>    try:<br>        proc = subprocess.Popen(cmd)<br>```<br><br>**Snyk analysis**<br><br>Unsanitized input from a command line argument flows into subprocess.Popen, where it is used as a shell command. This may result in a Command Injection vulnerability. | **GPT-4 analysis**<br><br>Language: Python<br><br>Vulnerabilities Detected:<br><br>Command Injection: Using subprocess.Popen() directly with user-controlled input can introduce command injection vulnerabilities, especially if cmd can be influenced externally.<br><br>Logging Sensitive Data: Logging the exact command being executed might expose sensitive data or internal workings that could be exploited if logs are not secured or inadvertently disclosed.<br><br>Incomplete Code: The provided code snippet appears to be incomplete. Without handling the process results or exceptions properly, there could be unexpected behavior or undisclosed errors.<br><br>Proposed Fixes:<br><br>Sanitize Input: Ensure that cmd is properly sanitized, especially if it can be influenced by external input. Prefer using the list format for the |
| | cmd argument in Popen which avoids shell interpretation.
Instead of: subprocess.Popen("ls -l")
Use: subprocess.Popen(["ls", "-l"])
Limit Logging: Consider logging only necessary information without exposing the exact command or arguments. If detailed logs are essential for debugging, ensure that the logs are protected and access-controlled.
Handle Process Result: Ensure you handle the result of the process. Capture stdout, stderr, and handle any potential exceptions from the Popen call. Consider using subprocess.run() for more straightforward command execution with better error handling. |
| **analyzer/python/ikos/analyzer.py**
```
# Fix huge slow down when ikos-analyzer
uses DROP TABLE on an existing db
    if os.path.isfile(db_path):
        os.remove(db_path)
```
**Snyk analysis**
Unsanitized input from a command line argument flows into os.remove, where it is used as a path. This may result in a Path Traversal vulnerability and allow an attacker to remove arbitrary files. | Language: Python

Vulnerabilities Detected:

Uncontrolled File Deletion: The script deletes a file without performing any checks about what the file might be. If an attacker can influence db_path, they might be able to delete arbitrary files the script has permissions to access.
Lack of Error Handling: The code doesn't handle potential exceptions that os.remove might raise if, for instance, the file is currently in use or if permissions prevent its deletion.
File Path Injection: If db_path is derived from user input or an untrusted source without validation, an attacker can inject arbitrary paths.
Proposed Fixes:

Validate File Path: Always validate file paths before performing actions |

| | |
|---|---|
| | on them. In this case, you might want to ensure that db_path is within a specific directory or matches a certain expected pattern. |
| | Error Handling: Wrap the file removal in a try-except block to handle potential exceptions gracefully. |
| | python |
| | Copy code |
| | try: |
| |   if os.path.isfile(db_path): |
| |     os.remove(db_path) |
| | except Exception as e: |
| |   print(f"Error while deleting file {db_path}: {e}") |
| | Limit Permissions: Run scripts like this with the least privilege required. If the script's primary purpose is to delete specific database files, then it shouldn't have permissions to delete other unrelated files. |
| | Code Fixes |
| | ```
    try:
        if os.path.isfile(db_path):
            os.remove(db_path)
    except Exception as e:
        print(f"Error while deleting file {db_path}: {e}")
``` |
| **Upload to GPT-4 Code Interpreter the public NGA file called radarsat.py [31]**<br>(1174 lines of python code)<br>https://github.com/ngageoint/sarpy/sarpy/io/complex/radarsat.py | **Code Interpreter Identification of Vulnerabilties:**<br>Based on the analysis of the visible part of the code, we have identified the following potential vulnerabilities:<br>Insecure File Operations:<br>Multiple uses of the open function. This can lead to vulnerabilities if user input is used directly to determine file paths without validation.<br>Insecure imports: |

| | |
|---|---|
| | Importing the os module. While the os module itself is not inherently insecure, certain functions within it (like os.system, os.popen, etc.) can introduce vulnerabilities if used improperly. |
| | Severity Ratings: |
| | Insecure File Operations: |
| | Severity: Medium. The severity can escalate to high if user-controlled input is used without validation for file operations. |
| | Insecure imports: |
| | Severity: Low. The mere import of the os module is not a direct vulnerability. However, if functions from the os module are misused, the risk can be much higher. |
| **Code Interpreter Identification of Fixes:**<br><br>1. Insecure File Operations:<br><br>The use of the open function can lead to vulnerabilities if user input is used directly to determine file paths without proper validation. This could allow an attacker to read arbitrary files (known as Path Traversal) or overwrite critical files.<br><br>Proposed Fixes:<br><br>Input Validation: Always validate user input if it's used to construct a file path. Ensure that the path is within the expected directory and doesn't contain sequences like ../ that could traverse to a different directory.<br><br>Use Absolute Paths: Instead of relative paths, use absolute paths wherever possible. This helps ensure that the application knows exactly where it's reading/writing data.<br><br>Error Handling: Implement proper error handling when opening files to avoid revealing sensitive system information.<br><br>2. Insecure Imports:<br><br>Importing the os module is not inherently insecure, but some functions within it can introduce vulnerabilities if used improperly.<br><br>Proposed Fixes:<br><br>Limit Use of the os Module: If you're only using a few functions from the os module, consider importing only those specific functions (e.g., from os import path). This can limit potential misuse.<br><br>Avoid Shell Commands: Avoid using functions that execute shell commands (e.g., os.system or os.popen) unless absolutely necessary. If you must use them, ensure that user input is not passed directly or is properly sanitized and validated before use. | |

| | |
|---|---|
| Use Safer Alternatives: If possible, use safer alternatives to os functions. For instance, use the subprocess module with a list of arguments instead of os.system to avoid shell injection vulnerabilities. | |